\documentclass[10pt,a4paper]{article}
\usepackage{geometry}
\usepackage{abstract}
\usepackage{adjustbox}
\usepackage{amsfonts,mathrsfs, amssymb, amsthm}
\usepackage{amsmath}
\usepackage{anysize}
\usepackage{array}
	\newcolumntype{L}[1]{>{\raggedright\let\newline\\\arraybackslash\hspace{0pt}}m{#1}}
	\newcolumntype{C}[1]{>{\centering\arraybackslash\hspace{0pt}}p{#1}}
\usepackage{authblk}
\usepackage[english]{babel}
\usepackage{booktabs}
\usepackage{calc}
\usepackage{caption}
\usepackage{changepage}
\usepackage{chngcntr}
\usepackage{color}
\usepackage{comment}
\usepackage{doi}
\usepackage{eurosym}
\usepackage{fancyhdr}
\usepackage{textcomp}
\usepackage{gensymb}
\usepackage{lastpage}
\usepackage{float}
\usepackage{graphicx}
\usepackage[]{hyperref}
\usepackage[utf8]{inputenc}
\usepackage{lipsum}
\usepackage{longtable}
\usepackage{multirow}
\usepackage{nicefrac}
\usepackage[intoc]{nomencl}
	\makenomenclature
	
\usepackage{paralist}
\usepackage{pbox}
\usepackage{pdflscape}
\usepackage{rotating}
\usepackage{setspace}
\usepackage{tabularx}
\usepackage{threeparttable}
\usepackage[nottoc]{tocbibind}
\usepackage[colorinlistoftodos]{todonotes}
\usepackage{url}
\usepackage{csquotes}
\usepackage{acronym}    
\usepackage{tabulary}
\usepackage{siunitx}

\usepackage{subfig}
\usepackage[subfigure]{tocloft}

\usepackage[sortcites, backend=biber, citestyle=numeric-comp, bibstyle=numeric, sorting=none, natbib]{biblatex}
\usepackage{enumitem} 
\usepackage{svg}

\marginsize{2.5cm}{2.5cm}{2cm}{2cm}
\allowdisplaybreaks

\title{An open tool for creating battery-electric vehicle time series from empirical data - \textit{emobpy}}

\author[]{Carlos Gaete-Morales}
\affil[]{\normalsize German Institute for Economic Research (DIW Berlin), Mohrenstr. 58, D-10117 Berlin, Germany, cgaete@diw.de.\vspace*{5pt}}
\author[]{Hendrik Kramer}
\affil[]{\normalsize Technische Universit\"at Berlin, D-10623 Berlin, Germany, hendrik.kramer@uni-due.de. \vspace*{5pt}}
\author[]{Wolf-Peter Schill}
\affil[]{\normalsize German Institute for Economic Research (DIW Berlin), Mohrenstr. 58, D-10117 Berlin, Germany, wschill@diw.de. \vspace*{5pt}}
\author[]{Alexander Zerrahn}
\affil[]{\normalsize German Institute for Economic Research (DIW Berlin), Mohrenstr. 58, D-10117 Berlin, Germany, azerrahn@diw.de.\vspace*{5pt}}
\date{\today}

\usepackage{parskip} 

\usepackage[compact]{titlesec}
\titlespacing{\section}{0pt}{*0}{*0}
\titlespacing{\subsection}{0pt}{*0}{*0}
\titlespacing{\subsubsection}{0pt}{*0}{*0}

\usepackage{etoolbox}
\apptocmd\normalsize{%
 \abovedisplayskip=6pt plus 3pt minus 9pt
 \abovedisplayshortskip=0pt plus 3pt
 \belowdisplayskip=6pt plus 3pt minus 9pt
 \belowdisplayshortskip=7pt plus 3pt minus 4pt
}{}{}

\begin{document}

\renewcommand{\thepage}{\roman{page}}
\maketitle
\thispagestyle{empty}
\thispagestyle{empty}

\begin{abstract}
There is substantial research interest in how future fleets of battery-electric vehicles will interact with the power sector. To this end, various types of energy models depend on meaningful input parameters, in particular time series of vehicle mobility, driving electricity consumption, grid availability, or grid electricity demand. As the availability of such data is highly limited, we introduce the open-source tool \textit{emobpy}. Based on mobility statistics, physical properties of vehicles, and other customizable assumptions, it derives time series data that can readily be used in a wide range of model applications. For an illustration, we create and characterize~$200$ battery-electric vehicle profiles for Germany. Depending on the hour of the day, a fleet of one million vehicles has a median grid availability between $5$ and $7$ gigawatts, as vehicles are parking most of the time. Four exemplary grid electricity demand time series illustrate the smoothing effect of balanced charging strategies.

\end{abstract}

\vfill

\newpage
\renewcommand{\thepage}{\arabic{page}}
\setcounter{page}{1}
\thispagestyle{fancy}


\section{Introduction}

We introduce \textit{emobpy}. It is an open-source python-based tool that creates profiles of battery-electric vehicles (BEV), based on empirical mobility statistics and customizable assumptions. We additionally provide a first application of the tool and create vehicle profiles based on representative German mobility data. An \textit{emobpy} profile consists of four time series: (i) \textit{vehicle mobility} containing the vehicle's location and distance travelled, (ii)~\textit{driving electricity consumption}, specifying how much electricity is taken from the battery for driving; (iii) \textit{BEV grid availability}, providing information whether and with which power rating a BEV is connected to the electricity grid at a certain point in time; and (iv) \textit{BEV grid electricity demand}, specifying the actual charging electricity drawn from the grid, based on different charging strategies. 

Such profiles are core input data for a wide range of model applications in energy, environmental, and economic studies on BEV. Technology developments as well as energy and climate policy measures drive the deployment of BEV in many countries~\citep{eia_2019}. Growing BEV fleets can have substantial impacts on the power sector. They increase the electric load, but may also provide temporal flexibility for integrating variable renewable energy sources and contribute to decarbonizing transportation~\citep{ipcc_2018}. Many model-based analyses investigate potential power sector interactions of future BEV fleets~\citep{Daina_2017a, mwasilu_2014, richardson_2013, Taljegard_2019} and thus depend on a meaningful representation of electric vehicles' mobility patterns. 

Yet such data are often not publicly available. In general, empirical data are scarce because BEV fleets are still small in most countries. And if respective time series are available, they are often specific to the conditions in which the data was collected and subject to data protection provisions. Past approaches make either stylized coarse assumptions~\citep{Kempton_2005}, derive data from mobility statistics, but lack documentation, transparency or reproducibility~\citep{Fischer_2019,Kiviluoma_2011,Muratori_2018,robinson_2013,schaeuble_2017,schill_2015,Heinz_2018}, or are idiosyncratic with respect to geographic characteristics or assumed driver behavior~\citep{Chen_2018,robinson_2013,Wolinetz_2018}. 

\begin{figure}[H]
\centering
\includegraphics[width=0.75\textwidth]{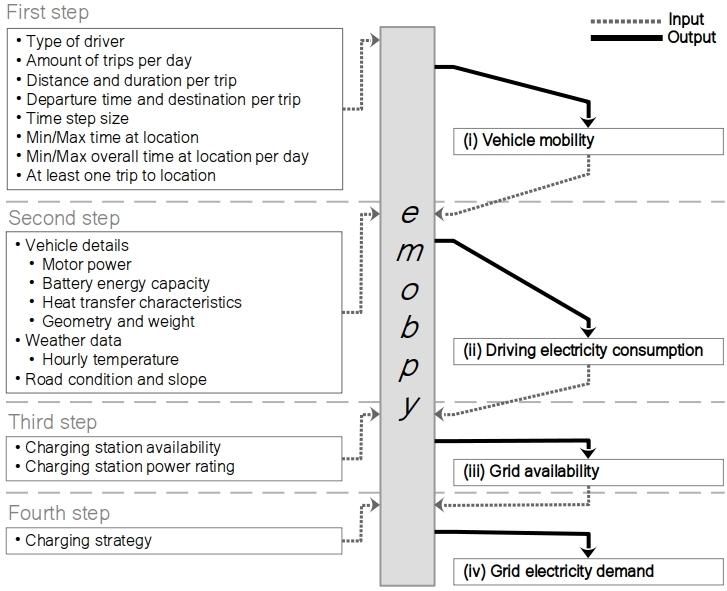}
\caption{Inputs and outputs for emobpy and sequence of generating four types of time series. The boxes on the left-hand side show customizable input assumptions, the boxes on the right-hand side indicate the four types of time series.}
\label{figure:scenario}
\end{figure}

Following~\citep{Daina_2017a}, we argue that new models are needed to derive relevant time series in a transparent and flexible way. As a first step in this direction, the tools Vencopy \citep{Wulff2020} and RAMP-mobility \citep{LOMBARDI2019} recently emerged. To further fill this gap, we developed \textit{emobpy}. Our tool takes empirical mobility statistics, physical properties of vehicles, and customizable assumptions as inputs and delivers BEV profiles as output. Figure~\ref{figure:scenario} gives a stylized account. We first discuss the outputs, then the required inputs. 

Four output time series constitute one BEV profile. These profiles have a customizable length and resolution. A handy format for many applications is all hours of one year. But also other formats are possible by discretion of the researcher. Likewise, the researcher can choose how many profiles she wants to create. 

The time series of vehicle mobility (i) contains the location of the vehicle at each time step and the time steps during which the vehicle is driving with information of the distance traveled. The driving electricity consumption time series~(ii) provides information on how much electricity the vehicle consumes for driving in each time step. The time series of (iii) grid availability provides information whether a vehicle is connected to the electricity grid in a time step and if so, with what power rating for charging or discharging. The time series of grid electricity demand (iv) provides information on how much electricity a vehicle demands from the electricity grid in a time step. Time series (i), (ii) and (iii) are core inputs for models that endogenously determine the timing of charging (and, potentially, discharging to the grid); the time series (iv) are core inputs for models that do not endogenously determine the grid interactions of BEV, but use exogenous input data for this.

The required input data for the time series of vehicle mobility~(i) are the relative frequencies of different driver types, e.g.,~commuters, of the number of trips per day, of the destination, distance and duration of trips, and of the departure hours. Such information can often be derived from national mobility statistics. If required or desired, a researcher can also make up own assumptions or resort to the pre-set values from German mobility statistics. \textit{emobpy} makes sure that the resulting time series are feasible and consistent. To this end, a minimum and maximum number of hours at specific locations can specified, and it is assumed that the last trip of a day heads home. With a Monte Carlo approach, \textit{emobpy} ensures variability across profiles. 

Based on the vehicle mobility time series, the driving electricity consumption~(ii) time series is derived. This requires further input data, such as information on nominal motor power, curb weight, drag coefficient, and dimensions, which the tool includes for several current BEV models. Ambient temperature is also a significant parameter that affects the consumption of BEV \citep{Brown_2018,Fischer_2019}. For that reason, \textit{emobpy} is endowed with a database of hourly temperature for European countries with a registry of the last 17 years. Additionally, the vehicle cabin insulation characteristics are required; this data is not widely available and thus assumed independently of the BEV models database. Driving cycles are also important input parameters that are used to simulate every individual trip. The model includes two driving cycles, \textit{Worldwide Harmonized Light Vehicles Test Cycle} (WLTC) and \textit{Environmental Protection Agency} (EPA). This input data is already provided within the tool, and the user can select a particular BEV model, country weather and driving cycle. Additionally, \textit{emobpy} also allows providing user-defined custom data.

The required input data for the grid availability time series~(iii) is the driving electricity consumption time series~(ii). Further, data or assumptions on the power rating of charging stations at different generic locations as well as their availability probabilities are needed. Variability across profiles is, again, introduced through a Monte Carlo approach, while \textit{emobpy} makes sure that the time series~(iii) is consistent within each profile .

The required input data for the grid electricity demand time series~(iv) includes the created time series on driving electricity consumption~(ii) and grid availability~(iii). Additionally, users can choose a charging strategy, such as immediate full charging or night-time charging, or make customary assumptions.


\section{Results}\label{sec:results}


\subsection{Application to Germany: parameterization and setup}\label{sec_applicationgermany}

For a first application of \textit{emobpy}, we draw on the comprehensive German mobility survey \textit{Mobilität in Deutschland}~\citep[Mobility in Germany,][]{MID_2017}. The survey features mobility data relating to different types of households, vehicles, individuals, and trips. In this application, we make three general assumptions: first, we assume that individuals with access to a vehicle carry out all their trips with the same vehicle; second, we assume that future BEV drivers have similar mobility patterns as current conventional drivers covered by the underlying mobility statistics; and third, for simplicity and tractability, we assume that there are only four BEV models: Hyundai Kona, Renault Zoe, Tesla Model~3 and Volkswagen ID.3. These models had the largest market shares in Germany by the time of writing. Again, all of these pre-set assumptions can easily be modified in \textit{emobpy}.

We generate~$50$ profiles for each BEV model, i.e.,~$200$ BEV profiles overall, each consisting of four time series. We focus on two types of drivers: commuters ($62$\% of all drivers) and non-commuters ($38$\% of all drivers). For commuters, we further differentiate between full-time and part-time employees, with a split of~$78$ to~$22$\%~\citep{OCDE_2018}. We exclude commuting students, apprentices, and trainees, who represent only a small share of all commuters in the initial dataset. The amount of trips per day varies between~$0$~-~$5$ with different probabilities for weekdays and weekend days (Table~\ref{tab:trips_number}).

\vspace{0.5cm}
\begin{table}[th!]
\caption{Probability distributions (given in~\%) for the amount of trips per day by days of the week}
\centering\label{tab:trips_number}
\begin{threeparttable}  
\begin{small}
\begin{tabular}{C{2.5cm}C{3.5cm}C{3.5cm}}
\toprule
Number of trips &     Working days & Weekend days  \\
\midrule
0        &        35.4 &       50.7  \\
1        &        0.0 &         0.0  \\
2        &       29.9 &        27.5  \\
3        &        8.3 &         4.4  \\
4        &       12.5 &        10.2  \\
5        &       13.9 &         7.2  \\
\bottomrule
\\ [-2ex]
\multicolumn{3}{p{0.65\textwidth}}{\textit{Note:} Data adapted from~\citep{MID_2017}. Commuters have the same distribution of daily trips as non-commuters. Data correspond to the group of respondents that have a yearly mileage in the range of $10,000$-$15,000$~km.
} \\
\end{tabular}
\end{small}
\end{threeparttable}
\end{table}
\renewcommand{\baselinestretch}{1.4}

The trip distance and duration follows a probability distribution derived from the input data (Table~\ref{tab:trips_distance}). As the underlying mobility statistics features a category that includes any trips with more than 100~km distance and more than 60~minutes duration, we cap the maximum distance travelled per trip at~$400$~km and the trip duration at~$185$ minutes. We also ensure that the average velocity resulting from every possible combination of distance and duration cannot exceed $130$~km/h.

\vspace{0.5cm}
\begin{table}[th!]
\caption{Joint probability distributions (given in~\%) for the distance travelled by trip and trip duration.}
\centering
\label{tab:trips_distance}
\begin{threeparttable}
\begin{small}
\begin{tabular}{C{2cm}C{1.2cm}C{1.2cm}C{1.2cm}C{1.2cm}C{1.2cm}C{1.2cm}C{1.2cm}}
\toprule
 & \multicolumn{7}{c}{Trip duration (minutes)} \\
 \cmidrule{2-8}
Distance & 10 & 10 - 15 & 15 - 20 & 20 - 30 & 30 - 45 & 45 - 60 & 60 - 185 \\
\midrule
1 km         &  2.9  &  0.3  &  0    & 0 & 0   & 0   & 0   \\
1 - 2 km     &  3.5  &  4.8  &  0.8  & 0 & 0   & 0   & 0   \\
2 - 5 km     &  8.4  & 10.2  &  5.7  & 0 & 1.2 & 0.4 & 0   \\
5 - 10 km    &  1.3  & 12.2  & 14.4  & 0 & 2.4 & 0.6 & 0.7 \\
10 - 20 km   &  0    &  0.9  &  6.3  & 0 & 4.7 & 1.3 & 0.5 \\
20 - 50 km   &  0    &  0    &  0    & 0 & 8.6 & 2.1 & 1.6 \\
50 - 100 km  &  0    &  0    &  0    & 0 & 0   & 0.6 & 2.1 \\
100 - 400 km &  0    &  0    &  0    & 0 & 0   & 0   & 1.5 \\
\bottomrule 
\\ [-2ex]
\multicolumn{8}{p{0.85\textwidth}}{\textit{Note:} Data adapted from~\citep{MID_2017}. Numbers rounded to one decimal. Data correspond to the group of respondents that have a yearly mileage in the range of $10,000$-$15,000$~km. All values add up to 100\%.
} \\
\end{tabular}
\end{small}
\end{threeparttable}
\end{table}
\renewcommand{\baselinestretch}{1.4}

The probability of departure times is specific to the trip destination, type of driver, and day of the week (Table~\ref{tab:trips_purpose_time}). It is distributed according to the underlying mobility statistics. Following the input data, we consider six trip destinations: \textit{workplace}, \textit{shopping}, \textit{errands}, \textit{escort}, \textit{leisure}, and \textit{home}. An example for errands is a visit to the doctor or to the authorities. In the case of escort destinations, the driver transports other persons, for example children. A set of rules is implemented in this case study to select only consistent day trips. The rules are applied depend on day of the week and the type of driver (Table~\ref{tab:rules} in the Methods section).

\vspace{0.5cm}
\begin{table}[th!]
\centering
\begin{scriptsize}
\caption{Joint probability distributions (given in~\%) for trip destinations and departure times, differentiated for commuters and non-commuters and days of the week}
\centering\label{tab:trips_purpose_time}
\begin{threeparttable}
\begin{tabular}{C{1.60cm}C{0.95cm}C{0.73cm}C{0.73cm}C{0.73cm}C{0.73cm}C{0.73cm}C{0.73cm}C{0.74cm}C{0.74cm}C{0.74cm}C{0.74cm}}
\toprule
 & Workplace & \multicolumn{2}{c}{Shopping} & \multicolumn{2}{c}{Errands} & \multicolumn{2}{c}{Escort} &  \multicolumn{2}{c}{Leisure} & \multicolumn{2}{c}{Home} \\
\cmidrule(l{0.5em}r{0.5em}){2-2} \cmidrule(l{0.5em}r{0.5em}){3-4} \cmidrule(l{0.5em}r{0.5em}){5-6} \cmidrule(l{0.5em}r{0.5em}){7-8} \cmidrule(l{0.5em}r{0.5em}){9-10} \cmidrule(l{0.5em}r{0.5em}){11-12}
Commuter & yes & yes & no & yes & no & yes & no & yes & no & yes & no \\ [-0.0ex]
\cmidrule(l{0.5em}r{0.5em}){2-2} \cmidrule(l{0.5em}r{0.5em}){3-4} \cmidrule(l{0.5em}r{0.5em}){5-6} \cmidrule(l{0.5em}r{0.5em}){7-8} \cmidrule(l{0.5em}r{0.5em}){9-10} \cmidrule(l{0.5em}r{0.5em}){11-12} \\
Departure & \multicolumn{11}{c}{\textbf{Working days}} \\ 
\cmidrule(l{0.5em}r{0.5em}){1-1} \cmidrule(l{0.5em}r{0.5em}){2-12}
\\ [-2ex]
05:00-08:00 & 11.1 & 0.5 & 0.7 & 0.5 & 0.7 & 1.1 & 0.7 & 0.5 & 0.7 & 0.8  & 0.7  \\
08:00-10:00 & 3.1  & 1.8 & 4.5 & 1.4 & 4.1 & 0.8 & 0.9 & 1.4 & 3.2 & 1.8  & 3.6  \\
10:00-13:00 & 1.3  & 2.7 & 6.7 & 2.3 & 5.4 & 0.7 & 1.3 & 3.2 & 4.7 & 5.5  & 11.7 \\
13:00-16:00 & 1.1  & 2.5 & 3.7 & 2.2 & 4.0 & 1.8 & 1.5 & 3.8 & 5.9 & 8.9  & 8.2  \\
16:00-19:00 & 0.3  & 3.0 & 1.9 & 2.2 & 2.2 & 1.4 & 1.0 & 4.9 & 4.5 & 14.0 & 9.3  \\
19:00-22:00 & 0.3  & 0.4 & 0.1 & 0.6 & 0.4 & 0.4 & 0.3 & 2.4 & 1.5 & 6.1  & 4.0  \\
22:00-05:00 & 0.6  & 0.0 & 0.0 & 0.1 & 0.1 & 0.1 & 0.1 & 0.4 & 0.2 & 2.4  & 1.3  \\
\\
 & \multicolumn{11}{c}{\textbf{Saturday}} \\  \cmidrule(l{0.5em}r{0.5em}){2-12}
\\ [-2ex]
05:00-08:00 & 0.9 & 1.2 & 1.2 & 0.3 & 0.3 & 0.2 & 0.2 & 0.8 & 0.8 & 0.8 & 0.8  \\
08:00-10:00 & 0.5 & 4.8 & 4.9 & 1.9 & 2.0 & 0.7 & 0.7 & 2.7 & 2.8 & 3.0 & 3.1  \\
10:00-13:00 & 0.4 & 7.1 & 7.3 & 3.5 & 3.6 & 1.4 & 1.5 & 5.2 & 5.4 & 9.1 & 9.3 \\
13:00-16:00 & 0.2 & 3.4 & 3.5 & 2.5 & 2.6 & 1.2 & 1.2 & 7.0 & 7.1 & 7.6 & 7.8  \\
16:00-19:00 & 0.1 & 2.3 & 2.4 & 1.7 & 1.7 & 1.1 & 1.1 & 6.0 & 6.1 & 9.5 & 9.7  \\
19:00-22:00 & 0.1 & 0.4 & 0.4 & 0.5 & 0.5 & 0.4 & 0.4 & 2.5 & 2.6 & 4.9 & 5.0  \\
22:00-05:00 & 0.2 & 0.0 & 0.0 & 0.1 & 0.1 & 0.2 & 0.2 & 0.8 & 0.8 & 3.0 & 3.1  \\
\\
 & \multicolumn{11}{c}{\textbf{Sunday}} \\  \cmidrule(l{0.5em}r{0.5em}){2-12}
\\ [-2ex]
05:00-08:00 & 0.8 & 0.3 & 0.3 & 0.2 & 0.2 & 0.1 & 0.1 & 0.8  & 0.8  & 0.4 & 0.4  \\
08:00-10:00 & 0.4 & 1.5 & 1.5 & 1.4 & 1.5 & 0.6 & 0.6 & 4.8  & 4.9  & 2.0 & 2.1  \\
10:00-13:00 & 0.3 & 0.7 & 0.7 & 2.8 & 2.8 & 1.3 & 1.3 & 11.7 & 11.9 & 7.2 & 7.4 \\
13:00-16:00 & 0.3 & 0.5 & 0.5 & 2.6 & 2.6 & 1.4 & 1.4 & 13.7 & 14.0 & 8.8 & 9.0  \\
16:00-19:00 & 0.2 & 0.2 & 0.2 & 1.8 & 1.9 & 1.0 & 1.0 & 6.8  & 7.0  & 13.3 & 13.6  \\
19:00-22:00 & 0.2 & 0.1 & 0.1 & 0.5 & 0.5 & 0.4 & 0.5 & 2.0  & 2.1  & 6.4 & 6.6  \\
22:00-05:00 & 0.1 & 0.0 & 0.0 & 0.1 & 0.1 & 0.1 & 0.1 & 0.4  & 0.4  & 2.0 & 2.1  \\
\bottomrule 
\\ [-2ex]
\multicolumn{11}{l}{\textit{Note:} Data adapted from~\citep{MID_2017}. Numbers rounded to one decimal.} \\
\end{tabular}
\end{threeparttable}
\end{scriptsize}
\end{table}
\renewcommand{\baselinestretch}{1.4}

Table~\ref{tab:ev_models} contains information on the four BEV models used for this case study. Most of the parameters serve to calculate driving electricity consumption, with the exception of nominal battery capacity that is used to generate the grid availability time series (iii)and grid demand time series (iv). Many other parameters are also provided by \textit{emobpy} to calculate the driving electricity consumption, such as efficiencies, auxiliary power and heat transfer data, as shown in Tables~\ref{tab:ev_insulationarray}~and~\ref{tab:consumption} in the Methods section. These are default values in the tool; however, they can be modified if desired by the user.

\vspace{0.5cm}
\begin{table}[th!]
\caption{BEV models' parameters derived from manufacturer data \citep{EVDB}}
\centering
\label{tab:ev_models}
\begin{threeparttable}
\begin{small}
\begin{tabular}{C{1.5cm}C{1.2cm}C{1.2cm}C{1.1cm}C{1.2cm}C{1.4cm}L{4.5cm}}
\toprule
 \multicolumn{2}{c}{ } & \multicolumn{4}{c}{BEV models} & \\
 \cmidrule{3-6}
Parameter & Unit & Model 3 (Tesla) & ID.3 (VW) & Kona (Hyundai) & Zoe (Renault) & \multicolumn{1}{c}{Description} \\
\midrule
$N_{motor}$  &  kW  &   358  &  93   & 150  & 65   & Nominal motor power \\
$N_{battery}$&  kWh &  79.5  &  45.0   & 64.0   & 45.6 & Nominal battery capacity \\
$m_{c}$      &  kg  &  1860  & 1600  & 1685 & 1480 & Curb weight  \\
$C_{d}$      &  -   &  0.23  & 0.27  & 0.29 & 0.29 & Drag coefficient  \\
$h$          &  m   &  1.44  & 1.55  & 1.57 & 1.56 & Height  \\
$w$          &  m   &  1.85  & 1.81  & 1.80 & 1.73 & Width  \\
$r_{gear}$   &  -   &  9.0     & 10.0    & 8.0    & 9.3  & Gear ratio \\
$PMR$        & W/kg &  192   & 58    & 89   &  44  & Power to mass ratio \\ 
\bottomrule 
\\ [-2ex]
\end{tabular}
\end{small}
\end{threeparttable}
\end{table}
\renewcommand{\baselinestretch}{1.4}

We model time steps of $15$ minutes. In each time step, a vehicle is either \textit{driving} in case a trip takes place, or is in one of the locations \textit{workplace}, \textit{shopping}, and so on. Depending on the vehicle location, a charging station to connect the vehicle to the grid may be available with a location-specific power rating. For this application, we assume four generic types of charging stations with different probability distributions for each vehicle location. The charging stations are at \textit{home}, in the \textit{public} area, or at the \textit{workplace}, or \textit{none} is available. Respective power ratings are~$3.6$, $22$, $11$, and~$0$~kW, based on~\citep{bmvi_2017}. The tool also considers fast charging; this feature is available for long-distance trips that are larger than the vehicle maximum range. The charging capacities selected for this application are $75$ and $150$ ~kW. This can be interpreted as the vehicle making a short stop during a longer trip. Charging efficiency is set to~$90$\%~(cp.~\citep{Kiviluoma_2011, Chen_2018, Taljegard_2019, Iora2019}).

When at \textit{home}, $81$\%~of all drivers park their vehicles in a carport or garage and~$19$\% on public streets according to~\citep{MID_2017}. For the group of vehicle profiles that have a carport at home, we assume a~$100$\% charging availability. For those without a private charging station, we set a probability of~$50$\% to find a \textit{public} charging station and~$50$\% of finding \textit{none}. For commuters, we consider three charging groups with different grid connection opportunities during work hours: charging at the \textit{workplace}, charging in the \textit{public} area, or \textit{none}. When commuters park their BEV at the \textit{workplace}, we assume that~$50$\% of them can charge their vehicles there, with a~$100$\% probability of finding a charging station; $25$\% of commuters charge in a \textit{public} area, with a~$50$\% probability of finding a charging station; and the remaining~$25$\% of commuters are assumed to have a~$100$\% probability of not having a charging station available during work hours (\textit{none}). For the vehicle locations \textit{shopping}, \textit{errands}, \textit{escort}, and \textit{leisure}, we assume a probability of~$50$\% to find a \textit{public} charging station and~$50$\% to find \textit{none}. When driving, grid connection is not available, with the exception of fast-charging for very long trips.

To derive time series of BEV grid electricity demand, we apply four exemplary charging strategies. Note that these charging strategies do not take into account any power sector or electricity market price information: 
\begin{itemize}
    \item \textit{immediate - full capacity}: BEV charge their batteries at full power rating as soon as they arrive at charging stations. Charging stops when the battery is full, or when the next trip starts. This mimics a setting where drivers have no incentives and/or no technical possibility to charge their vehicle batteries in a more balanced way, which is likely to be sub-optimal with respect to the electricity market or network situation.
    \item \textit{immediate - balanced}: BEV start charging their batteries as soon as they arrive at charging stations, however with constant power rating (usually below the power rating of the charging station), such that a~$100$\% state of charge is reached just before starting the next trip, assuming perfect foresight of the next departure time. This approximates a smoother and potentially more system-oriented charging behaviour. 
    \item \textit{at home - balanced}: similar to the previous charging strategy, but BEV only charge \textit{at~home}, even when additional charging options are available at other locations. This reflects a preference or economic incentive for home charging.
    \item \textit{at home night-time - balanced}: similar to the previous charging strategy, but with charging time restricted to the time window between~23:00 and~8:00. This mimics the effect of potential tariff incentives for night-time (off-peak) charging.
\end{itemize}

\subsection{Vehicle mobility}\label{sec_vehstates}

\begin{figure}[h!]
\centering
\includegraphics[width=0.99\textwidth]{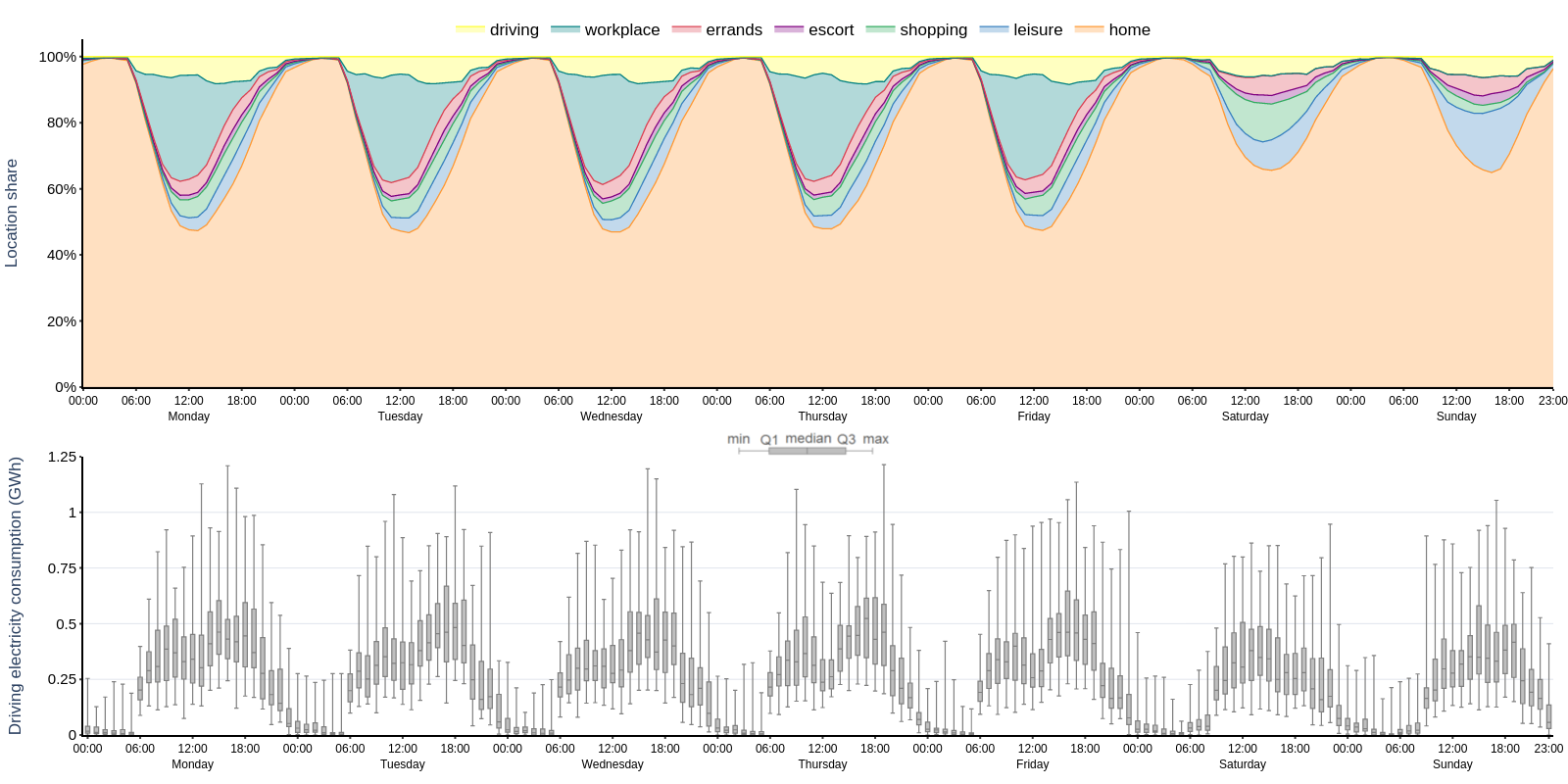}
\caption{Simulated time series of vehicle locations (top panel) and driving electricity consumption (bottom panel) of one million BEV, given as averages and box plots for each hour of the week.}
\label{fig:driving_result}
\end{figure}

Figure~\ref{fig:driving_result} summarizes all~$200$ simulated vehicle mobility time series. For each hour, vehicle locations are averaged over all profiles and weeks of the year. Hourly driving electricity consumption is summarized in box plots, rendering the dispersion over the simulated profiles through the weeks of the year. All numbers are linearly scaled up to represent one million BEV, so the setting may be interpreted as a near-term future scenario.

Most of the time, vehicles are parking (top panel). At night, between~23:00 and~5:00, more than~$96$\% of the fleet are, on average, at \textit{home}. During daytime, still the majority of vehicles are at \textit{home}, but also a large proportion of vehicles is at the \textit{workplace}, peaking at~$32$\% at~11:00 on working days. During weekends, more vehicles stay at \textit{home}, and the shares of \textit{shopping}, \textit{errands}, \textit{escort}, and \textit{leisure} increase. Commuters have a positive but very small probability of going to the workplace on weekends (Tables~\ref{tab:trips_purpose_time}~and~\ref{tab:rules}), so it is hardy visible in Figure~\ref{fig:driving_result}. Every day between~6:00 and~22:00, at least~$3$\% of the fleet are \textit{driving}, with a peak between around~15:00 and~17:00 with about~$9$\% of the fleet \textit{driving}.

\begin{figure}[h!]
\centering
\includegraphics[width=0.7\textwidth]{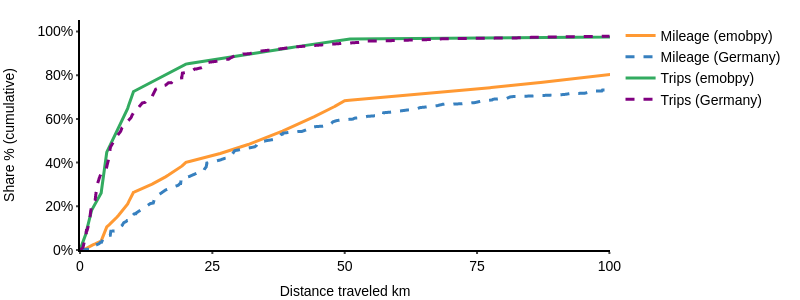}
\caption{Comparison of cumulative shares of trips and mileage per distance travelled. ``Germany'' represents German mobility statistics \citep{MID_2017}, which reports these aggregate shares up to a distance of~$100$~km.}
\label{fig:distance_validation}
\end{figure}

To validate \textit{emobpy} results, we compare the cumulative distributions of trips and mileage to the underlying German mobility statistics~\citep{MID_2017}. For the two metrics, the cumulative distributions follow a similar pattern (Figure~\ref{fig:distance_validation}). Both our \textit{emobpy} application and the official German statistics indicate that about~$90$\% of all trips have a distance travelled of~$50$~km or below. The cumulative mileage -- the overall distance travelled by all vehicles in a year -- also has a similar shape in \textit{emobpy} and in the official statistics up to~$40$~km. The Figure also allows inferring that long-distance trips above 100 km represent $25$\% of the yearly mileage, while those trips only account for $3$\% of all trips \citep{MID_2017}


\subsection{Driving electricity consumption}\label{sec_driving_cons}

The overall hourly driving electricity consumption of one million BEV (Figure~\ref{fig:driving_result}, bottom panel) peaks between~15:00 and 17:00 on working days with an annual median around $450$~MWh, and an absolute maximum at~19:00 of~$1250$~MWh. During the weekend, overall consumption is lower and with less distinctive evening peaks.

\begin{figure}[h]
\centering
\includegraphics[width=0.99\textwidth]{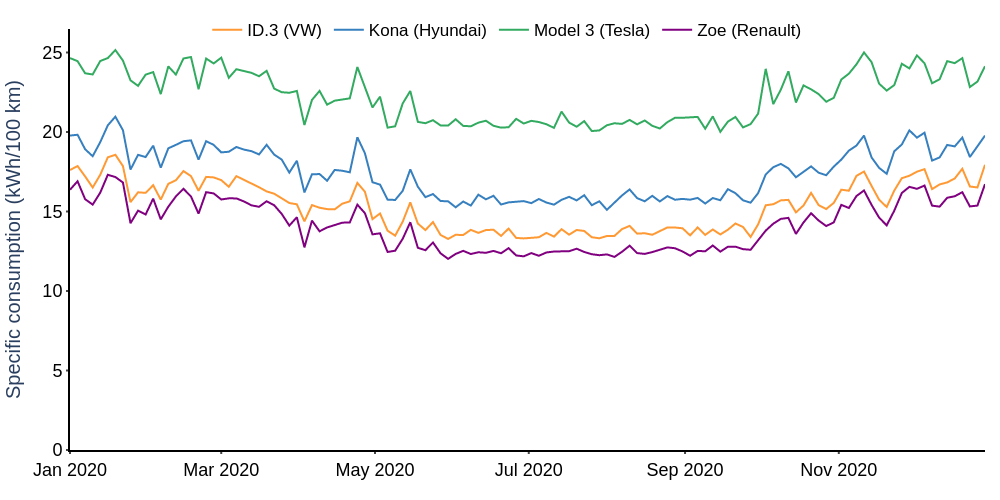}
\caption{Specific consumption of four selected BEV models throughout a year. The values are calculated as the medians of all trips taken for every three days.}
\label{fig:specific_consumption}
\end{figure}

The average specific consumption for all trips over the 200 profiles and four models is $22.2$~kWh/100~km. The median specific consumption values for the different BEV types Model~3, Kona, ID.3 and Zoe are $22.7$, $17.9$, $15.7$, and $14.5$~kWh/100~km, respectively (Figure \ref{fig:specific_consumption}). The ambient temperature variation has a clear impact on the specific consumption. On average, specific consumption is lowest in summer with $20.7$~kWh/100~km, and highest in winter with $23.6$~kWh/100~km.

\subsection{Grid availability}\label{sec_gridavail}

\begin{figure}[h]
\centering
\includegraphics[width=0.99\textwidth]{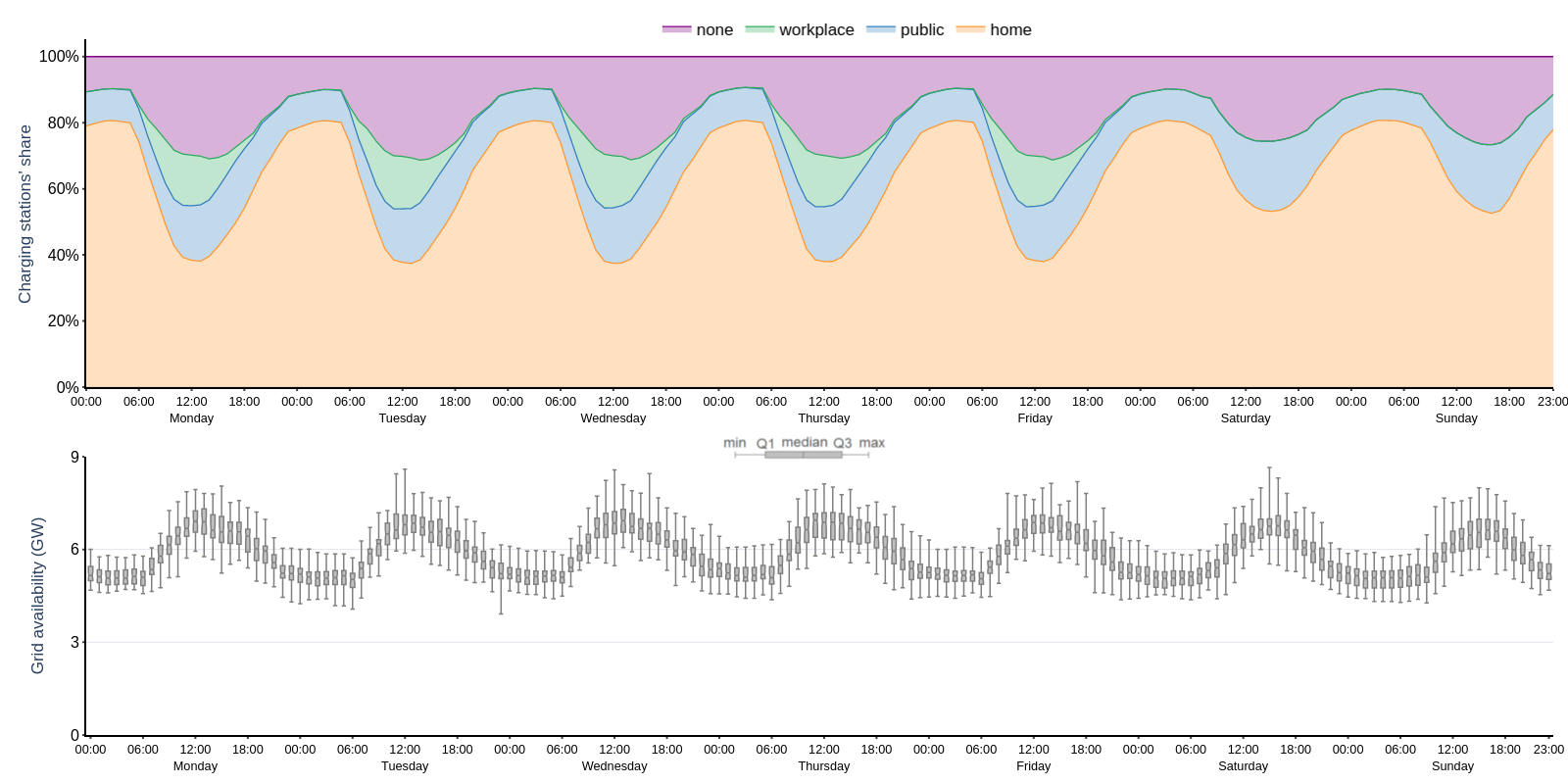}
\caption{Simulated time series summarized for different types of charging stations (top panel) and grid-connected power rating (bottom panel) of one million BEV, given as averages and box plots for each hour of the week.}
\label{fig:availability_result}
\end{figure}

The cumulative simulated grid availability time series is shown in Figure~\ref{fig:availability_result}. On working days, the time series on the types of charging stations has a recurring pattern (top panel) that corresponds to the pattern of vehicle locations. The share of vehicles with a charging station available reaches a~$90$\%~peak between~3:00 and~5:00 at night. Here, around~$80$\% of vehicles are connected at \textit{home} and~$10$\% on a \textit{public} street. Between~11:00 and~12:00, average grid availability is at a minimum level of~$70$\%. During daytime, a relevant proportion of available charging stations is at the \textit{workplace}. On weekends, the charging station time series is less peaky, with higher proportions at \textit{home} and on \textit{public} streets during daytime.

The grid-connected power rating is lowest between~19:00 and~8:00, with a median between~$5.0$ and~$5.6$~GW for a fleet of one million BEV (bottom panel). This is due to the high share of home charging stations with a low power rating of~$3.6$~kW. During daytime, the median grid-connected power rating is greater than~$7$~GW because charging stations available either at the \textit{workplace} or in \textit{public} areas have a power rating of~$11$ and~$22$~kW, respectively.

\subsection{Grid electricity demand}\label{sec_chargepatt}

\begin{figure}[h]
\centering
\includegraphics[width=0.99\textwidth]{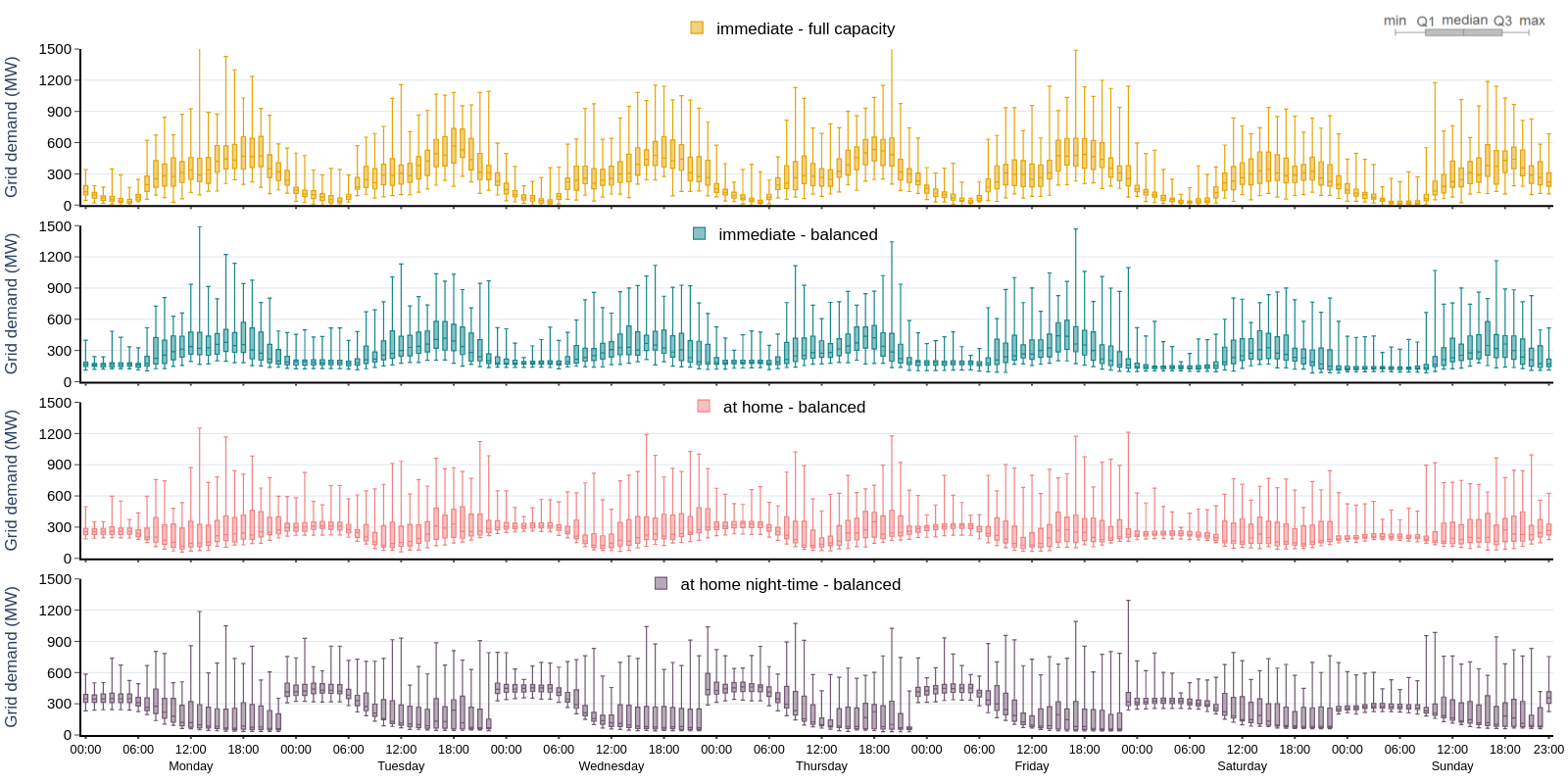}
\caption{Simulated grid electricity demand time series for a fleet of one million BEV for four charging strategies, summarized in box plots for each hour of the week.}
\label{fig:actualcharge_result}
\end{figure}

The grid electricity demand time series for the four exemplary charging strategies are summarized in Figure~\ref{fig:actualcharge_result}. The \textit{immediate - full capacity} charging strategy leads to a volatile cumulative BEV grid electricity demand both over the week and over the year, with a pronounced diurnal pattern. A distinctive peak of hourly electricity demand from the grid, with median values around~$460$~MWh for a fleet of one million BEV, occurs on working day afternoons between~17:00 and~20:00, when many vehicles arrive at \textit{home} and charge immediately at full power rating. As the entire BEV fleet is assumed to charge similarly in this scenario, such a charging strategy would substantially add to the evening peak of electric load. It could thus have substantial repercussions on the power sector and other electricity consumers. Load peaks could increase even further if higher power rating for charging at home beyond~$22$~kW was considered.

The \textit{immediate - balanced} and \textit{at home - balanced} charging strategies have smoother temporal grid electricity demand patterns with lower peaks, because vehicles do not get charged at full power rating once they reach a charging station. Both the variance of medians and (interquartile) ranges are lower. Likewise, the median hourly consumption of the one million BEV fleet rarely exceeds~$400$~MWh for \textit{immediate - balanced}, and $300$~MWh for \textit{at home - balanced}. During weekdays, fluctuations are more pronounced for \textit{at home - balanced}, as most vehicles are at \textit{home} every night. Compared to \textit{immediate - full capacity}, such smoother charging may be more compatible with the power sector.

The \textit{at home night-time - balanced} charging strategy shows a distinct load peak at working day nights, with median hourly grid electricity demand of one million BEV around~$420$~MWh. Between Friday evening and Monday morning, median demand at night-time is lower than $300$~MWh because the vehicles are less used on weekends than on working days. Accordingly, any regulatory measures that shift BEV charging to night-time periods would lead to substantially less smooth patterns compared to all-day charging. Yet the power sector implications of these charging strategies are less clear and should be investigated in detail with dedicated energy models.


\section{Discussion}\label{sec_discussion}

The open-source tool \textit{emobpy} allows to derive electric mobility time series from empirical mobility data in a transparent and customizable way. The central outputs are profiles for individual BEV, consisting of four basic types of quarter-hourly time series covering a full year: vehicle mobility, driving electricity consumption, grid availability, and grid electricity demand. The number of vehicle profiles can be freely chosen. A greater number of profiles represents a large and diverse BEV fleet more realistically, yet may lead to greater computational burden when using the time series in energy model applications. Users may customize the tool and alter both the German mobility data used here and the various assumptions we made, such as the shares of driver types or the availability and power rating of charging stations.

The generated vehicle profiles can be used as inputs for a wide range of model analyses of electrified and decarbonized mobility futures. Research questions in energy, environmental, and economic studies requiring temporally detailed data of BEV are abundant. These comprise the role of BEV as flexibility resource to make efficient use of renewable electricity, emission effects of electric mobility, the impact of new loads from BEV on electricity prices, or electricity market repercussions of optimized versus user-driven charging schedules. 

Several limitations offer scope for future research. First, the object of study in \textit{emobpy} is the vehicle. Addressing the individual choice of the modal split would be an interesting complementary approach. This would also allow to relax the assumption that all trips are made with the same vehicle. Second, \textit{emobpy} draws on past mobility behavior data that does not necessarily reflect future behavior. While this is a generic issue in \textit{ex-ante} analyses, the model is flexible to accommodate alternative assumptions on future or counterfactual scenarios. Third, using input data on the distance and destination of trips, \textit{emobpy} determines vehicle locations as background information for creating a BEV profile. While this is a convenient approach to simulate temporal variation, it has no explicit spatial resolution. We argue that this is a minor drawback because many energy, environmental, and economic models rather address a macro perspective without zooming into fine spatial detail. Further, we exclude a group of drivers that have a \textit{service} trip destination according to \citep{MID_2017}. This refers to profiles with numerous work-related trips per day, e.g.,~taxi drivers, which is conceptually challenging to model in our current framework. As we publish the code open-source under a permissive license, we expect that future and potentially collaborative development could address these points.

\section{Methods}\label{sec_methods}

One BEV profile consists of four time series: (i) vehicle mobility, (ii) driving electricity consumption, (iii) grid availability, and (iv) grid electricity demand. Time series (i) is created first. All the following time series will build up from this time series as it has locations at every time step and distance travelled while driving. Then, time series (ii) is calculated, taking time series (i) as an input. Time series (iii) is created, based on time series (ii); and time series (iv) is generated taking into account (ii) and (iii). For this Methods section, we introduce the following definitions: 

\begin{itemize}
    \item Edge: link or vertex that connects two nodes, where each node comprises an origin or a destination of a trip.
    \item Trip: edge with departure time, distance travelled and duration of the travel as attributes.
    \item Tour: also referred to as a day tour, it consists of a list of chronologically sorted trips by departure time. A tour contains all trips carried out by a BEV in a day.
\end{itemize}

The sampling approach consists of a sampling procedure of discrete choices. Input parameters are discrete choices with given corresponding probabilities~\citep[see][]{Numpy_2011}. Additionally, and only for the sampling of distance-duration-relations of trips, a second sampling is carried out if the probability distribution contains discrete distance ranges and duration ranges. In this case, a uniform distribution of integers is assumed to obtain a distance value which is within the distance range. The duration of the trip is subsequently obtained by interpolating the sampled distance with the respective distance range and duration range (see Table~\ref{tab:trips_distance}).


\subsection{Vehicle mobility}\label{sec:methods_subsec:mobility}

\begin{figure}[tb!]
\centering
\includegraphics[width=0.81\textwidth]{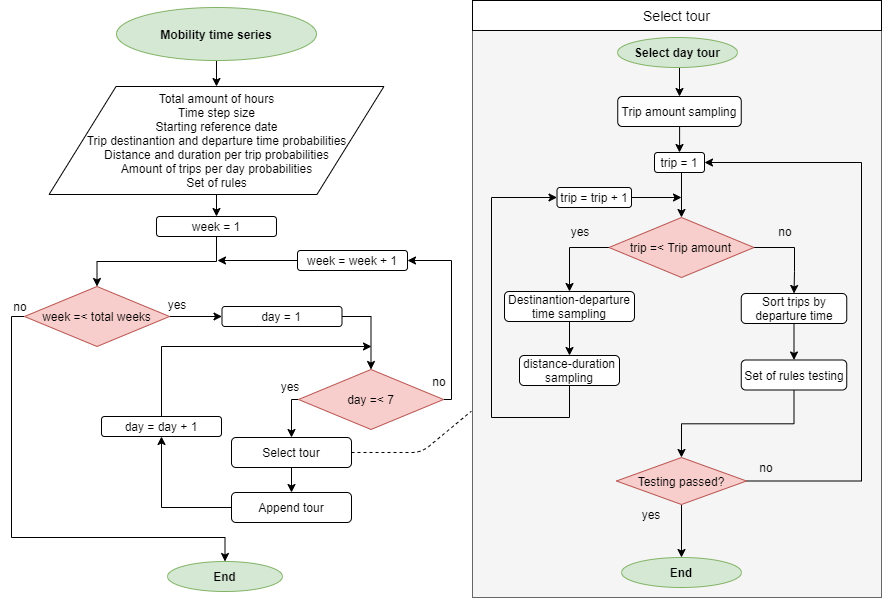}
\caption{Vehicle mobility time series flow diagram}
\label{fig_driving}
\end{figure}

The flow diagram shown in Figure~\ref{fig_driving} illustrates how \textit{emobpy} creates the time series of vehicle mobility. The input data are shown in the parallelogram in the left panel. The proportion of commuters and non-commuters is based on empirical data or assumptions. Additionally, the total time frame as a number of total weeks must be specified. A reference date can be used to map the day of the week. This is not only useful when the input statistics differentiates between weekdays, but also for allocating the temperature, which is a step required in the creation of time series (ii). Further inputs are three probability distributions that contain the number of day trips, the destinations and departure times, and the distance-duration-relation of the trips (compare Tables~\ref{tab:trips_number},~\ref{tab:trips_distance},~\ref{tab:trips_purpose_time}). Finally, a set of rules ensures that the tours are plausible (compare Table~\ref{tab:rules}).

The function~\textit{Select Tour} creates a plausible day tour. Its output is a chronologically sorted list of trips, where each trip is represented by an edge of two locations (origin and destination) with departure time, distance travelled, and trip duration. This function is used twice as displayed in the left-hand side of Figure~\ref{fig_driving}.

For every day of the calculation period, the function \textit{Select Tour} is called. Initially, a number of trips for the current day is obtained by sampling from the probability distribution that matches the type of driver. Trips are sampled according to the joint probability distribution of destinations and departure times. The sampled trips are stored in a sequential order. For each new sampled trip, \textit{emobpy} disregards all tuples that contain the departure time of the already selected tuples, and the probability of the remaining tuples is normalized to add up to~$100$\%. This avoids selecting a destination-departure time tuple with the same departure time as the one already selected. Once the total amount of tuples matches the number of daily trips, the sampling is finished and the tuples are ordered chronologically.

From the chronologically ordered tuples of destination and departure time, the eventual trips are created by establishing an origin-destination edge with its departure time as an attribute. The distance travelled and trip duration for each trip is sampled from the probability distribution provided by distance-duration statistics, such as shown in Table~\ref{tab:trips_distance}. Distance and duration of each trip are also attached to the origin-destination edge as attributes.

The duration time at each location is calculated from the arrival time and departure time. The arrival time is estimated from the previous trip departure time and trip duration. The next step evaluates the feasibility of the tour by checking the set of rules (Table~\ref{tab:rules}), such as the minimum time at the workplace or whether the last trip heads home. All rules must be satisfied, or the current tour is discarded and the process is repeated until feasible results are obtained. 

\begin{table}[th!]
\caption{Rules implemented to select consistent day trips}
\resizebox{\textwidth}{!}{%
\begin{tabular}{L{4.3cm}L{2.8cm}C{2cm}C{1.5cm}C{2cm}C{1.5cm}C{2cm}C{1.5cm}}
\toprule
\multicolumn{2}{r}{} & \multicolumn{2}{c}{Non-commuter} & \multicolumn{2}{c}{Full-time commuter} & \multicolumn{2}{c}{Part-time commuter} \\
\cmidrule{3-8}
\multicolumn{2}{c}{Rule} & Working day & Weekend & Working day & Weekend & Working day & Weekend\\
\midrule
Minimum time at & home               & 0.5 hrs & 0.5 hrs & 0.5 hrs & 0.5 hrs & 0.5 hrs & 0.5 hrs \\
                & workplace          & - & - & 3.5 hrs & 3.0 hrs & 3.5 hrs & 3.0 hrs \\
                & other destinations & 0.5 hrs & 0.5 hrs & 0.5 hrs & 0.5 hrs & 0.5 hrs & 0.5 hrs \\
\midrule
Minimum time per day at & home               & 9 hrs & 6 hrs & 9 hrs & 6 hrs & 9 hrs & 6 hrs \\
                        & workplace          & - & - & 7 hrs & 3 hrs & 3.5 hrs & 3 hrs  \\
                        & other destinations & - & - & - & - & - & - \\
\midrule
Maximum time per day at & home               & - & - & - & - & - & - \\
                        & workplace          & - & - & 8 hrs & 4 hrs & 4 hrs & 4 hrs \\
                        & other destinations & - & - & - & - & - & - \\
\midrule
At least one trip to & home               & yes & yes & yes & yes & yes & yes \\
                     & workplace          & - & - & yes & no & yes & no \\
\bottomrule
\\ [-2ex]
\end{tabular}}
\label{tab:rules}
\end{table}

\subsection{Driving electricity consumption}\label{sec:methods_subsec:driving_cons}

\begin{figure}[H]
\centering
\includegraphics[width=0.65\textwidth]{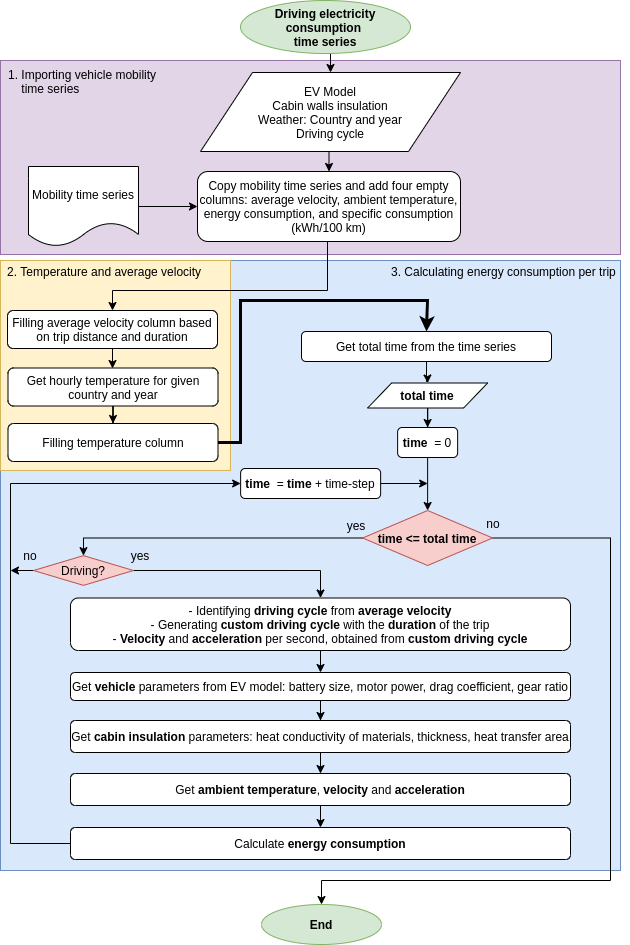}
\caption{Driving electricity consumption flow diagram}
\label{fig_consumption_flow}
\end{figure}

The flow diagram displayed in Figure~\ref{fig_consumption_flow} illustrates how \textit{emobpy} creates the driving electricity consumption time series. The first block describes the input data, including the vehicle mobility time series. Different types of input parameters are required. Parameters associated with the vehicle can be obtained by selecting a BEV model. This includes nominal motor power, the battery energy capacity, the curb weight, the drag coefficient, height and width to calculate the frontal area, the gear ratio, and power-to-mass ratio (compare Table~\ref{tab:ev_models}). Also, we make additional parameter assumptions associated with vehicles, such as battery charging and discharging efficiency, transmission system efficiency, cabin air volume, coefficient of performance of heat pumps and accessories' average power. The tool also requires passenger-related parameters, such as average weight and sensible heat, and the average number of passengers. Ambient temperature as well as driving cycle assumptions are also required. \textit{emobpy} has access to three types of datasets: a) Hourly temperature time series can be obtained for $39$ European countries \citep{DeFelice2018}; b)~Parameters of $25$ BEV models that can be retrieved from \cite{EVDB}; and c) cabin thermal insulation based on \cite{Wirth2013}. The default values used in our case study are defined in Tables \ref{tab:ev_insulationarray} and \ref{tab:consumption}. The second block consists of incorporating the temperature time series. Trip distance and duration are used to calculate the average velocity for each trip. The third block shows the steps for calculating the energy consumption for each trip. The respective trip average velocity and trip duration are used to generate a custom driving cycle from a standard driving cycle sub-class. In doing so, velocity and acceleration are simulated at high-resolution (per seconds), which enables us to calculate power flow and energy consumption as described in the following sections.

\begin{figure}[H]
\centering
\includegraphics[width=0.85\textwidth]{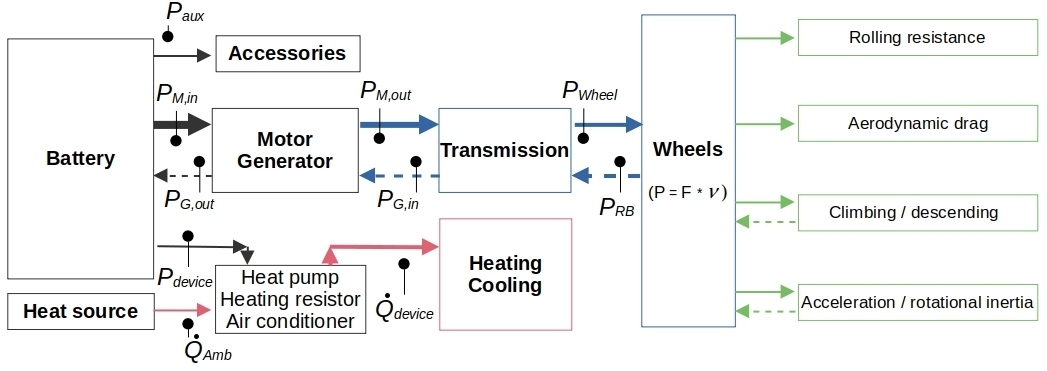}
\caption{Block diagram of the power flows at the components of the electric vehicle while driving. $P$: power, $F$: forces, $\nu$: velocity, $P_{aux}$: auxiliary power, $P_{M,in}$: motor input power, $P_{M, out}$: motor output power, $P_{device}$: electrical power for heating/cooling devices, $\dot{Q}_{Amb}$: heat transfer rate from ambient by heat pump, $\dot{Q}_{device}$: heat transfer rate for heating/cooling, $P_{G,in}$: generator input power, $P_{G,out}$: generator output power, $P_{Wheel}$: power at wheels, $P_{RB}$: regenerative braking power. [Black lines: electrical power, blue: mechanical power, red: heat transfer rate, green: acting forces. Dashed lines represent flows related to regenerative braking. Line thickness indicates typical flow magnitudes.]}

\label{fig_driving_power}
\end{figure}

To calculate a trip's energy consumption, we calculate the power requirements for vehicle traction, heating and cooling. We further include (customizable) assumptions on accessory power. Figure~\ref{fig_driving_power} shows the power flows between the battery and the wheels, the heating/cooling devices and the  accessories. 

\subsubsection{Custom driving cycles}\label{sec:methods_subsec:drivingCons_sub:cycles}
 
A custom driving cycle is required to simulate a vehicle's driving pattern based on the trip average velocity and trip duration. This is necessary to calculate the power flow of a vehicle journey. The \textit{Worldwide Harmonized Light Vehicles Test Cycle} (WLTC) is the tool's default driving cycle. A driving cycle emulates driving velocity patterns in cities, suburban areas, or highways, represented by driving cycle sub-classes. Every driving cycle sub-class has an average velocity which is calculated, including stops. The tool first selects the sub-class, whose average velocity is closest to the current trip's average velocity. The driving cycle sub-class selected is divided by the average velocity of the sub-class and multiplied by the trip's average velocity to create a custom driving cycle. This approach modifies the original driving cycle only to a small extent. Finally, as driving cycles have a finite duration, the custom driving cycle is replicated sequentially until the total driving cycle length reaches the trip duration. Acceleration is calculated from the variation of the velocity.

\subsubsection{Vehicle tractive effort}\label{sec:methods_subsec:drivingCons_sub:traction}

Tractive effort $F_{te}$ is the force required to surpass the opposing forces to the movement of a vehicle, expressed in Eq.~\ref{eq:1}, where $F_{rr}$ is the rolling resistance force, $F_{ad}$ is the aerodynamic drag, $F_{g}$ is the climbing force and $F_{acc}$ is the linear acceleration and inertia force \citep{GENIKOMSAKIS2017}.

\begin{align}
F_{te} &= F_{ad} + F_{rr} + F_{g} + F_{acc} \label{eq:1}\\
F_{ad} &= \frac{1}{2} \rho \cdot A_{frontal} \cdot C_{d} \cdot \nu^{2} \label{eq:2}\\
F_{rr} &= f_{rr} \cdot m \cdot g \cdot \cos \theta \label{eq:3}\\
F_{g} &= m \cdot g \cdot sin \theta \label{eq:4}\\
F_{acc} &= (m + m_{i}) \alpha \label{eq:5}
\end{align}

The aerodynamic drag force, as defined in Eq.~\ref{eq:2}, depends on $\rho$ moist air density, $A_{frontal}$ frontal area of the vehicle, $C_{d}$ drag coefficient, and $\nu$ vehicle's velocity. The rolling resistance force is displayed in Eq.~\ref{eq:3}, where $f_{rr}$~is the rolling resistance coefficient, $m$~is the vehicle mass, $g$~is the gravitational acceleration, and $\theta$~is the slope in radians. Climbing force is shown in Eq.~\ref{eq:4}, and linear acceleration and inertia force of rotating parts is presented in Eq.~\ref{eq:5} where $\alpha$ is linear acceleration, and $m_{i}$ is the inertial mass, a mass that represents the inertia of moving parts \citep{Iora2019}. The inertial mass is defined in Eq.~\ref{eq:8} that depends on the curb mass of the vehicle $m_{c}$ and the gear ratio $r_{gear}$, while the mass of the vehicle $m$ is the sum of the curb mass $m_{c}$ and the passengers mass $m_{p}$ as shown in Eq.~\ref{eq:7}. $f_{rr}$ is a parameter that depends on the ambient temperature $T_{amb}$ and velocity according to Eq.~\ref{eq:6}. This equation is derived from empirical data according to \cite{wevj2015}.

\begin{align}
m &= m_{c} + m_{p}  \label{eq:7}\\
m_{i} &= m_{c} (0.04 + 0.0025 r_{gear})  \label{eq:8} \\
f_{rr} &= \num{1.9e-6} T_{amb}^{2} - \num{2.1e-4} T_{amb} + 0.013 + \num{5.4e-5} \nu  \label{eq:6}
\end{align}

\subsubsection{Motor power}\label{sec:methods_subsec:drivingCons_sub:motor}

Power at wheels $P_{Wheel}$ is estimated at each time step, as shown in Eq.~\ref{eq:9} where $F_{te}$ is non-negative. Otherwise, $P_{Wheel}$ is zero and regenerative braking power takes the absolute value of $F_{te}$ as shown in Section \ref{sec:methods_subsec:drivingCons_sub:breaking}. The output power of the motor $P_{M,out}$ is defined in Eq.~\ref{eq:10} where $\eta_{tr}$ is the transmission system efficiency. The input power of the motor $P_{M,in}$ depends on its output power and the motor efficiency $\eta_{m}$, as shown in Eq.~\ref{eq:11}.

\begin{align}
P_{Wheel} &= F_{te} \cdot \nu \quad if~F_{te} > 0  \label{eq:9}\\
P_{M,out} &= \frac{P_{Wheel}}{\eta_{tr}}  \label{eq:10}\\
P_{M,in} &= \frac{P_{M,out}}{\eta_{m}}  \label{eq:11}
\end{align}

The motor efficiency $\eta_{m}$ depends on the motor's angular speed and torque. This value can be determined experimentally for each vehicle model or can be provided by the manufacturer. We have implemented a more general approach described in \cite{GENIKOMSAKIS2017} and  \cite{Iora2019} (Eq.~\ref{eq:13}). The efficiency function depends on the motor load fraction $Load_{m}$ as defined in Eq. \ref{eq:12} where $N_{motor}$ is the nominal power capacity of the motor.

\begin{align}
Load_{m} &= \frac{P_{M,out}}{N_{motor}}  \label{eq:12}\\
\eta_{m} &= f(Load_{m})  \label{eq:13}
\end{align}

\subsubsection{Regenerative braking} \label{sec:methods_subsec:drivingCons_sub:breaking}

Regenerative braking power $P_{RB}$ occur when $F_{te}$ is negative as defined in Eq. \ref{eq:14} where the absolute values of $F_{te}$ is used.
\begin{align}
P_{RB} &= \big| F_{te} \big| \cdot \nu \quad if~F_{te} < 0  \label{eq:14}
\end{align}

The input power of the generator $P_{G,in}$ is described in Eq.~\ref{eq:15} where $\eta_{rb}$ is the efficiency of the regenerative braking. The regenerative braking efficiency represents the fraction of the regenerative braking power that can be effectively recovered. The Eq.~\ref{eq:17} shows the regenerative braking efficiency is a function of the acceleration $\alpha$ \citep{FIORI2016}.
\begin{align}
P_{G,in} &= P_{RB} \cdot \eta_{tr} \cdot \eta_{rb}  \label{eq:15}\\
\eta_{rb} &= \bigg[ e^{\frac{0.0411}{\left| \alpha \right|}} \bigg]^{-1}  \label{eq:17} \\
P_{G,out} &= P_{G,in} \cdot \eta_{g}  \label{eq:16}
\end{align}
Assuming a generation efficiency $\eta_{g}$, we can estimate the output power of the generator $P_{G,out}$ as indicated in Eq.~\ref{eq:16}. The load fraction of the generator $Load_{g}$ is required to calculate the $\eta_{g}$ as shown in Eq.~\ref{eq:18}, where $N_{g}$ is the nominal power capacity of the generator that is in fact also the nominal power capacity of the motor. The dataset with corresponding $\eta_{g}$ by $Load_{g}$ is obtained from \cite{GENIKOMSAKIS2017} and  \cite{Iora2019} (see Eq.~\ref{eq:19}).

\begin{align}
Load_{g} &= \frac{P_{G,in}}{N_{gen}}  \label{eq:18}\\
\eta_{g} &= f(Load_{g})  \label{eq:19}
\end{align}

\subsubsection{Heating, cooling and accessories}\label{sec:methods_subsec:drivingCons_sub:hvac}

We aim to estimate the power that an electric device has to provide for heating or cooling a vehicle cabin to keep the temperature on a level of comfort for the passengers. To do so, we use a heat balance model \citep{shibata2015, klemm2018}. The heat balance equation is shown in Eq.~\ref{eq:20}. The left-hand side expression represents the amount of heat accumulated in the cabin air, where $V_{cabin}$ is the cabin volume, $\rho_{air, T_{cabin}}$ is moist air density at cabin temperature, $C_{p}$ is the specific heat of air, $T_{cabin}$ is the cabin temperature, and $\frac{dT_{cabin}}{dt}$ is the temperature change in the cabin over time. The right-hand side expression of the heat balance considers the following mechanisms: a) enthalpy of outside air $\dot{Q}_{inflow}$, b) enthalpy of discharged air to outside $\dot{Q}_{outflow}$, c) heat transfer through the cabin walls $\dot{Q}_{wall}$, d) sensible heat of passengers $\dot{Q}_{person}$, and e) the heat provided by a device to keep the target temperature in the cabin $\dot{Q}_{device}$. The device may be either a resistor or a heat pump. Radiation heat transfer and latent heat by condensation/evaporation are features not considered in this model.


\begin{align}
V_{cabin} \cdot \rho_{air, T_{cabin}} \cdot C_{p} \frac{dT_{cabin}}{dt} &= \dot{Q}_{device} + \dot{Q}_{person} + \dot{Q}_{inflow} - \dot{Q}_{outflow} - \dot{Q}_{wall}  \label{eq:20}\\
\dot{Q}_{inflow} &= \rho_{air,T_{amb}} \cdot \dot{V}_{in} \cdot C_{p} \cdot T_{amb}  \label{eq:22}\\
\dot{Q}_{outflow} &= \rho_{air,T_{cabin}} \cdot \dot{V}_{out} \cdot C_{p} \cdot T_{cabin}  \label{eq:23}\\
\dot{Q}_{wall} &= (T_{cabin} - T_{amb}) \sum_{k=1}^{n}\frac{1}{R_{k}}  \label{eq:24} \\
R_{k} &= \frac{1}{A_{k}}(\frac{1}{h_{cabin}} + \sum_{j=1}^{m} \frac{x_{j}}{\lambda_{j}} + \frac{1}{h_{amb}})  \label{eq:25} \\
h_{cabin} &= constant  \label{eq:26} \\
h_{amb} &=  \label{eq:27}
    \begin{cases}
      6.14 \nu^{0.78}, & \text{if}\ \nu > 5~m/s \\
      6.14 \cdot 5^{0.78}, & \text{otherwise}
    \end{cases}\\
\dot{Q}_{person} &= q_{sensible} \cdot n_{p}  \label{eq:21} \\
P_{device} &= \frac{\big| \dot{Q}_{device} \big|}{COP}  \label{eq:28} \\
P_{aux} &= constant  \label{eq:30}
\end{align}
The enthalpy of outside air $\dot{Q}_{inflow}$ is described in Eq.~\ref{eq:22}, where $\rho_{air,T_{amb}}$ is the moist air density in the ambient, $\dot{V}_{in}$ is the volume inflow of air for ventilation, and $T_{amb}$ is the ambient temperature. The enthalpy of discharged air to outside $\dot{Q}_{outflow}$ is defined in Eq.~\ref{eq:23}, where $\dot{V}_{out}$ is the output volume flow of air. The heat transfer through the cabin walls $\dot{Q}_{wall}$ is shown in Eq.~\ref{eq:24}, where $R_{k}$ is the heat transfer resistance and $k$ is the set of cabin zones. The heat transfer resistances $R_{k}$ is defined in Eq.~\ref{eq:25}, where $A_{k}$ is the area of every cabin zone,  $h_{cabin}$ is the convection heat transfer coefficient between the cabin air and the vehicle wall, $h_{amb}$ is the convection heat transfer coefficient between the wall and ambient air, $x_{j}$ is the thickness of thermal insulation material of the wall, $\lambda_{j}$ is the thermal conductivity, and $j$ is the set of insulation materials. The cabin convection heat transfer coefficient $h_{cabin}$ is defined in Eq.~\ref{eq:26} where it has been assumed a constant value. Typical values are $10$-$20$ $\frac{W}{m^{2}K}$ \citep{klemm2018}.

The ambient convection heat transfer coefficient $h_{amb}$ is defined in Eq.~\ref{eq:27}, where $\nu$ is the outside wind speed, which we consider to be equal to the vehicle's velocity \citep{FUJITA2001}. The sensible heat of passengers $\dot{Q}_{person}$ is presented in Eq.~\ref{eq:21}, where $q_{sensible}$ is the sensible heat per person and $n_{p}$ is the number of passengers. The heat balance equation is solved for $\dot{Q}_{device}$ to get the heat requirement. The electric power for the heating/cooling $P_{device}$ is defined in Eq.~\ref{eq:28}, where $COP$ is the coefficient of performance of the heater/cooler or heat pump. A constant power for accessories $P_{aux}$ is assumed as shown in Eq.~\ref{eq:30}. To estimate the heat transfer that occurs by heat conduction, Table~\ref{tab:ev_insulationarray} displays the default insulation configuration used in \textit{emobpy}.

\vspace{0.5cm}
\begin{table}[ht!]
\caption{Configuration of the vehicle cabin insulation \citep{Wirth2013,klemm2018,breque2017,Rashid2018}}
\centering
\label{tab:ev_insulationarray}
\begin{threeparttable}
\begin{small}
\begin{tabular}{C{0.5cm}L{2cm}C{1.53cm}C{1.45cm}C{0.85cm}C{0.6cm}C{1.3cm}C{1.45cm}C{1.0cm}}
\toprule
    &   & \multicolumn{6}{c}{Layers [\textit{j}]} &  Area\\
    \cmidrule{3-8}
    &   & Laminated glass & Tempered glass  & Metal & PU foam  & Polyester & Fiberglass  & ($m^{2}$) [$A_{k}$] \\
\midrule
\multirow{4}{*}{\rotatebox[origin=c]{90}{Zones [\textit{k}]}} & Windshield & \checkmark &  & & & & & 1.7  \\
    & Side windows  &  & \checkmark  &   &  &  &  & 1.5  \\
    & Rear window    &    & \checkmark &  &   &   &   & 1.4   \\
    & Rest & & & \checkmark & \checkmark & \checkmark & \checkmark & 9.9 \\
    \cmidrule{1-9}
\multicolumn{2}{l}{Thermal conductivity ($\frac{W}{m K}$) [$\lambda_{j}$]} & 0.6 & 1.38 & 60 & 0.02 & 0.64 & 2 & \\
\multicolumn{2}{l}{Layer thickness (mm) [$x_{j}$]} & 4.5  & 3.5  & 0.9  & 58 & 2  & 1 & \\
\bottomrule 
\\ [-2ex]
\multicolumn{9}{p{0.65\textwidth}}{\textit{Note:} PU:~Polyurethane.} \\
\end{tabular}
\end{small}
\end{threeparttable}
\end{table}
\renewcommand{\baselinestretch}{1.4}

\subsubsection{Energy consumption}\label{sec:methods_subsec:drivingCons_sub:battery}

Positive or negative values can be expected at the battery $P_{battery}$. Suppose the sum of motor input power, generator output power, auxiliary power and power for heating/cooling $P_{all}$ is positive (see Eq.~\ref{eq:31}). The battery then provides energy to the vehicle as it discharges (see Eq.~\ref{eq:32}) and the discharging efficiency is used $\eta_{discharge}$. If $P_{all}$ is negative, then the battery is charged via regenerative braking. In such a case, the battery load $P_{battery}$ is negative, hence the charging efficiency $\eta_{charge}$ is utilized. 

\begin{align}
P_{all} &= P_{M,in} - P_{G,out} + P_{aux} + P_{device}  \label{eq:31}\\
P_{battery} &=  \label{eq:32}
    \begin{cases}
      \frac{P_{all}}{\eta_{discharge}}, & \text{if}\ P_{all} > 0 \\
      P_{all} \cdot \eta_{charge}, & \text{otherwise}
    \end{cases}
\end{align}

\begin{align}
E_{total} &= \sum_{t=1}^{\tau} P_{battery,t} \label{eq:33}
\end{align}

The total energy consumption per trip $E_{total}$ is defined in Eq.~\ref{eq:33}, where battery load is aggregated through the set $t$ that consists of the duration of the trip at every second. Table~\ref{tab:consumption} provides parameters required for estimating the driving electricity consumption. For reasons of simplicity and data availability, we assume that these parameter do not differ between BEV models.

\vspace{0.5cm}
\begin{table}[th!]
\caption{Parameters used for all BEV models to determine driving electricity consumption}
\renewcommand{\arraystretch}{2}
\centering\label{tab:consumption}
\begin{threeparttable}
\begin{small}
\begin{tabular}{C{1.5cm}C{1.4cm}C{1.5cm}L{7cm}C{2cm}}
\toprule
Parameter &     Unit & Value & Description & Reference \\
\midrule
$\eta_{tr}$ &  \%  & 95 & Transmission efficiency & \citep{FIORI2016,GENIKOMSAKIS2017,Iora2019} \\
$\eta_{charge}$ & \% & 90 & Efficiency for battery charging & \citep{Kiviluoma_2011, Chen_2018, Taljegard_2019, Iora2019}\\
$\eta_{discharge}$ & \% & 95 & Efficiency for battery discharging & \citep{ZHANG2019}\\
$h_{cabin}$     & $\frac{W}{m^{2} K}$ & 10 & \parbox{7cm}{Cabin air convective heat transfer coefficient} & \citep{klemm2018, Rashid2018} \\
$m_{p}$     & $kg$ & 75 & Person mass & Own assumption \\
$q_{sensible}$ & $W$ & 70 & \parbox{7cm}{Person sensible heat of a driver or passengers} & \citep{TORREGROSAJAIME2015,breque2017} \\
$n_{p}$     & $quantity$ & 1.5 & Number of passengers in the vehicle & \citep{bundestag2018} \\
$T_{amb}$   & $Celsius$ &  Germany (2016) & \parbox{7cm}{Time series hourly temperature}  & \citep{DeFelice2018} \\
$T_{cabin}$ & $Celsius$ & 17 (20) & \parbox{7cm}{Target cabin temperature for heating (cooling)} & Own assumption \\
$V_{cabin}$ & $m^{3}$ & 3.5 & Air volume of vehicle's cabin & \citep{ZHANG2019} \\
$\dot{V}_{in}$,$\dot{V}_{out}$  & $\frac{m^{3}}{s}$ & 0.02 & \parbox{7cm}{Input (output) air flow at ambient (cabin) temperature (for ventilation)} & \citep{leong2010,ott2008} \\
$P_{aux}$   & $W$ & 300 & \parbox{7cm}{Auxiliary power for electronic accessories and battery heating} & \citep{GENIKOMSAKIS2017,Iora2019} \\
Driving cycle &  -  & WLTC  & \parbox{7cm}{Driving cycles} & \citep{ciuffo2015} \\
$COP$       & - & 2 & \parbox{7cm}{Coefficient of performance. Values $>1$ imply the use of a heat pump; similar COP for heating and cooling assumed} & \citep{BELLOCCHI2018,Cuevas2019} \\
\bottomrule
\\ [-2ex]
\end{tabular}
\end{small}
\end{threeparttable}
\end{table}
\renewcommand{\baselinestretch}{1.4}

\subsection{Grid availability time series}\label{sec:methods_subsec:gridavailability}

\begin{figure}[tb]
\centering
\includegraphics[width=0.7\textwidth]{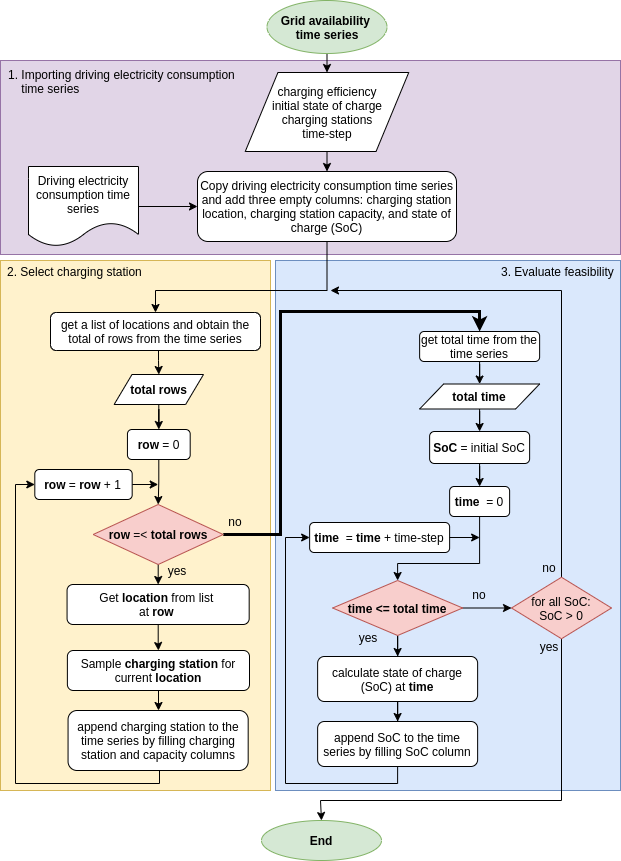}
\caption{Grid availability time series flow diagram}
\label{fig_grid}
\end{figure}

The flow diagram shown in Figure~\ref{fig_grid} illustrates how \textit{emobpy} creates the grid availability time series. Inputs are the time series of driving electricity consumption as well as locations and distances created in step~1. Further, \textit{emobpy} requires data or assumptions on the battery size, charging efficiency, the initial state of charge (SoC), and the probability distributions of charging stations at different locations including their respective power rating, as indicated in the parallelogram in the first box of Figure~\ref{fig_grid}.

Initially, a time series containing the time step, location, distance, and consumption columns is imported from the driving electricity consumption time series. Next, different types of charging stations are selected for each time step. For each parking location (arrival time plus subsequent parking time steps until next trip), the types and respective power ratings are sampled from the corresponding probability distributions.

After a candidate grid availability time series is created, \textit{emobpy} evaluates its feasibility. This check takes into account the driving electricity consumption time series of the profile as well as the charging station power rating available in each time step. To this end, the SoC of the battery is calculated for each time step by adding the energy taken from the grid for charging if connected to the grid, or subtracting the energy consumed from the battery if driving. For the first time step, we use an exogenous value. To simulate the SoC of the battery, we assume a charging strategy called \textit{immediate - full capacity} as introduced in Section~\ref{sec_chargepatt}. It draws electricity from the grid at full rating of the charging station as soon as the BEV is connected and until the battery is full. Section~\ref{sec:methods_subsec:griddemand} provides more detailed information. After calculating the SoC for all time steps, \textit{emobpy} verifies if each SoC lies within~$0$~-~$100$\%. If this is the case, the allocation of charging stations throughout the time series horizon allows to create a grid availability time series. If this is not the case, a new allocation is carried out. In case of many unsuccessful allocations, \textit{emobpy} returns a warning. Reasons comprise a low availability of charging stations and/or low power ratings compared to trip lengths or a low battery capacity.


\subsection{Grid electricity demand time series}\label{sec:methods_subsec:griddemand}

\begin{figure}[H]
\centering
\includegraphics[width=0.85\textwidth]{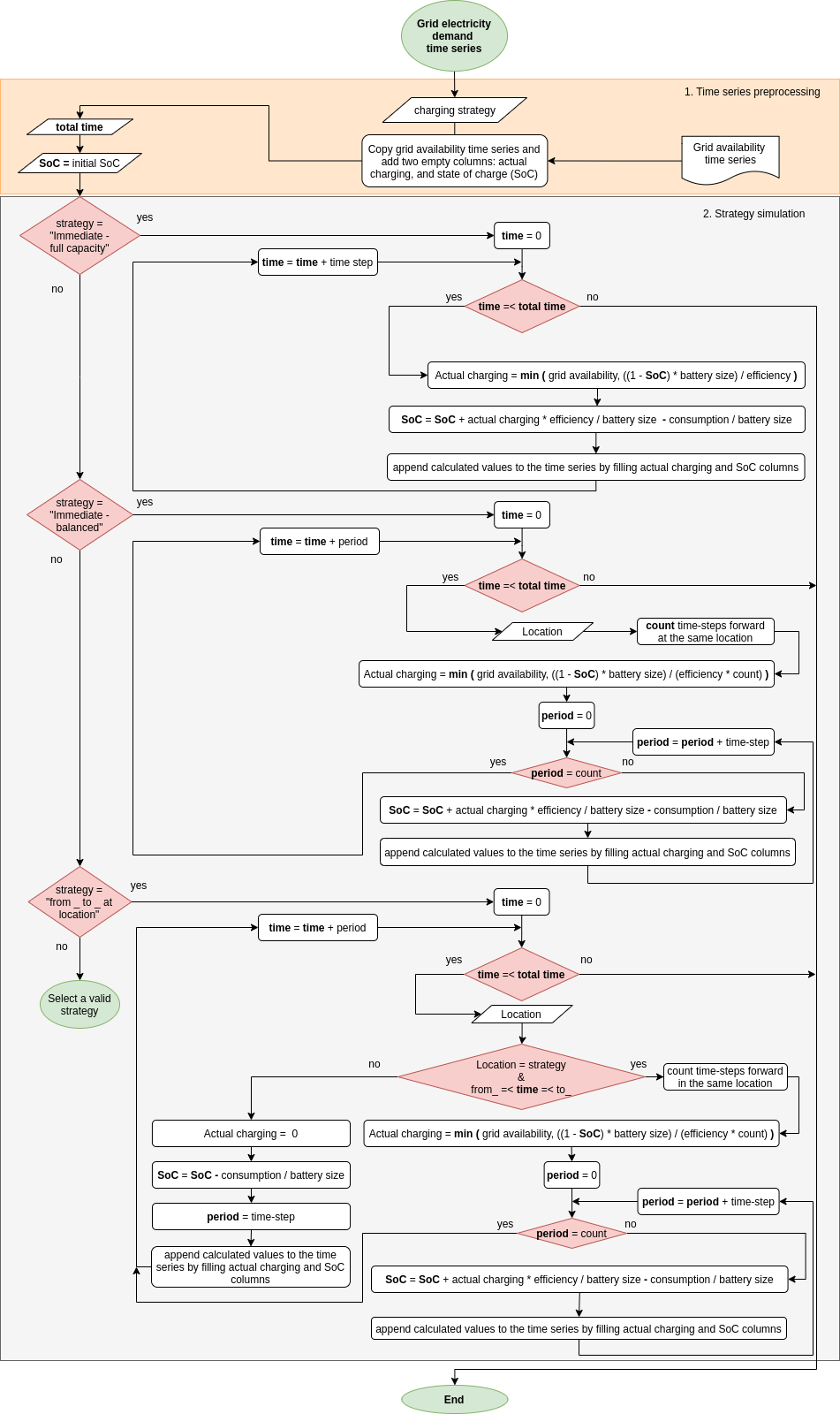}
\caption{Grid electricity demand time series flow diagram for the charging strategies \textit{immediate - full capacity}, \textit{immediate - balanced}, and a customized charging strategy}
\label{fig_demand}
\end{figure}

Figure~\ref{fig_demand} shows a flow diagram of how \textit{emobpy} creates the grid electricity demand times series. The inputs are the grid availability time series, including the charging power rating, and the driving electricity consumption time series, including vehicle locations. Further inputs are data or assumptions on the battery size, initial SoC, and charging efficiency. Based on the inputs, \textit{emobpy} calculates, for each time step, the SoC and, as output, the actual charging that represents the electricity drawn from the grid to charge the battery. To this end, two pre-set charging strategies (\textit{immediate - full capacity}, \textit{immediate - balanced}) or a customized charging strategy can be applied.

In the first pre-processing stage, \textit{emobpy} imports the input data. In the second stage, charging -- depending on the pre-set strategy -- and the according SoC of the battery are determined. 

For the strategy \textit{immediate - full capacity}, \textit{emobpy} iterates over all time steps without any foresight. It aims at reaching $100$\%~SoC as fast as possible. If the current time step indicates grid availability, this strategy charges the BEV at the full power rating, except when less than the full rate is required to obtain~$100$\% SoC. If the current time step corresponds to driving, actual charging is zero and the electricity consumed by the motor is subtracted from the SoC. 

For the strategy \textit{immediate - balanced}, \textit{emobpy} also charges the BEV as soon as a grid connection is available. Yet, based on perfect foresight, the model executes its iteration over all consecutive time steps a vehicle is parking at the same location. To this end, the energy required to fill up the battery completely ($100$\% SoC) is determined, and the resulting value is divided by the number of time steps that the vehicle remains parked. The actual charging equals the maximum station power rating only in case a~$100$\% SoC cannot be reached before the next trip. Otherwise, the actual charging rating is lower than the charging station power rating.

A customized charging strategy allows to derive alternative grid electricity demand time series. Such a strategy is passed to the model as text, e.g.,~\textit{From\_23\_to\_06\_at\_home}. In this example, the actual charging occurs in the time window defined in hours of the day~($23$~-~$06$) and when the vehicle is parked at a predetermined location~(\textit{home}). The charging is performed in \textit{balanced} configuration as described above. If a negative SoC is identified in the time series, the model may charge the battery outside the boundary defined by the customized charging strategy.


\section*{Data availability}

The dataset generated during the current study is available in the Zenodo repository \url{https://doi.org/10.5281/zenodo.3931663}~\citep{Gaete2021b}.


\section*{Code availability statement}

The tool can be installed from the Python Package Index (PyPI) at \url{https://pypi.org/project/emobpy/}. The code is provided under a permissive license in Zenodo \citep{Gaete2021}. We also provide the script created to generate the 200 BEV profiles for the current case study at \url{https://gitlab.com/diw-evu/emobpy/emobpy\_examples}.


\section*{Author contributions}
Conceptualization, W.P.S. and A.Z.; Methodology, C.G., H.K., W.P.S., and A.Z.; Software, C.G.; Investigation, C.G.; Writing - original draft, C.G., H.K., W.P.S, and A.Z.; Writing - review \& editing, C.G. and W.P.S.; Visualization, C.G.; Project administration and funding acquisition, W.P.S.


\newpage

\section*{References}
\printbibliography[heading=none]


\section*{Acknowledgements}
We thank three anonymous reviewers for valuable comments. We acknowledge a research grant of the German Federal Ministry of Education and Research (BMBF) via the START project (FKZ~03EK3046). Wolf-Peter Schill carried out parts of his work on this paper during a research stay at the Climate and Energy College at the University of Melbourne. The authors also thank all colleagues of the Open Energy Modelling Initiative for providing feedback during the Open Energy Modelling Workshop - Berlin 2020.


\section*{Competing interests\label{sec: declaration}}
The authors declare no competing interests.

\end{document}